\documentclass[conference]{IEEEtran} 
\pdfoutput=1
\IEEEoverridecommandlockouts
\usepackage[T1]{fontenc}
\usepackage{ifpdf}
\usepackage{algorithmic,graphicx,booktabs,textcomp,xcolor,color,url,hyperref,amsmath,cleveref,tabularx,balance,ifthen,amssymb,multirow,mdframed,soul}
\usepackage[inline]{enumitem}
\usepackage{flushend} %for balanced columns on last page
\usepackage[numbers]{natbib}
\newlist{ExpCase}{enumerate}{1}
\setlist[ExpCase]{resume,leftmargin=1.60cm,labelindent=\parindent,label=\textbf{S-\arabic*}:}

 \makeatletter

 \def\ps@IEEEtitlepagestyle{%
   \def\@oddfoot{\mycopyrightnotice}%
   \def\@evenfoot{}%
 }
 \def\mycopyrightnotice{%
   {\begin{minipage}{\textwidth}
   \footnotesize \copyright 2021 IEEE. Personal use of this material is permitted. Permission from IEEE must be obtained for all other uses, in any current or future media, including reprinting\slash republishing this material for advertising or promotional purposes, creating new collective works, for resale or redistribution to servers or lists, or reuse of any copyrighted component of this work in other works.
   \end{minipage}
   }% <--- Change here
   \gdef\mycopyrightnotice{}% just in case
 }

\begin{document}

\title{Cases for Explainable Software Systems: Characteristics and Examples}
\author{\IEEEauthorblockN{Mersedeh Sadeghi}
\IEEEauthorblockA{Software and Systems Engineering\\ University of Cologne\\
Cologne, Germany \\
sadeghi@cs.uni-koeln.de}
\and
\IEEEauthorblockN{Verena Klös}
\IEEEauthorblockA{Software and Embedded Systems Engineering\\Technische Universität Berlin\\
Berlin, Germany\\
verena.kloes@tu-berlin.de}
\and
\IEEEauthorblockN{Andreas
Vogelsang}
\IEEEauthorblockA{Software and Systems Engineering\\ University of Cologne\\
Cologne, Germany \\
vogelsang@cs.uni-koeln.de}}

\maketitle

\begin{abstract}
% Context
The need for systems to explain behavior to users has become more evident with the rise of complex technology like machine learning or self-adaptation. 
% Problem
In general, the need for an explanation arises when the behavior of a system does not match the user's expectations. However, there may be several reasons for a mismatch including errors, goal conflicts, or multi-agent interference. 
Given the various situations, we need precise and agreed descriptions of explanation needs as well as benchmarks to align research on explainable systems.  
% Contribution
In this paper, we present a taxonomy that structures needs for an explanation according to different reasons. We focus on explanations to improve the user interaction with the system. For each leaf node in the taxonomy, we provide a scenario that describes a concrete situation in which a software system should provide an explanation. These scenarios, called explanation cases, illustrate the different demands for explanations.
% Conclusions/Impact
Our taxonomy can guide the requirements elicitation for explanation capabilities of interactive intelligent systems and our explanation cases build the basis for a common benchmark. We are convinced that both, the taxonomy and the explanation cases, help the community to align future research on explainable systems. 
\end{abstract}

\section{Introduction\label{sec:intro}}
Today, the necessities and advantages of explainable intelligent systems are not a subject of doubt. Many studies have shown that explanations % explainability capability 
allow users to comprehend the behavior of an intelligent system such as robots \cite{stange2019towards}, interpret machine learning models \cite{paez2019pragmatic}, understand the reasons behind the decisions of an autonomous agent or recommender system, and more confidently determine how to proceed with such decisions and interact with intelligent agents~\cite{ tintarev2011designing}. 
That ultimately increases the trust of the user %and intelligent system
\cite{dovsilovic2018explainable, lyons2017shaping} and raises the chances of acceptance and utilization of such a system ~\cite{papamichail2003explaining}.

Research on explainable systems in computer science has evolved in different directions, broadly categorized as first, technical approaches towards the development of tools and techniques in a wide variety of domains, including but not limited to artificial intelligence, robotics, cyber-physical systems, and decision-making systems~\cite{samek2019towards, stange2019towards, blumreiter2019towards, goud2008development,vago2021visualization}. Second, empirical studies on the foundations of explanations. However, the latter constitutes only the minority of the works \cite{adadi2018peeking, langer2021we, nunes2017systematic}. 
Among them, most of the research has been oriented towards concepts such as assessment of the explainable system and their impacts in various domains \cite{bussone2015role, bunt2012explanations,kulesza2013too,narayanan2018humans}, explanation types, models and presentation formats \cite{jentzsch2019conversational,chromik2019dark, haynes2009designs,garcia2018explain,chari2020directions} and explanation and requirement engineering \cite{chazette2020explainability,vogelsang19}. Evidently, little attention has been paid towards the characterization of system behavior that confuses users and requires explanations. %what needs explanation.\\

To achieve a better understanding of when users can profit from explanations, we provide a taxonomy of such situations and give examples from the domain of cyber-physical systems in a smart home environment. We call these examples \textit{explanation cases} and envision them to build the basis for a benchmark of scenarios that can be used in research on explainable systems. Besides some more obvious cases, where the system behaves differently than requested, we also provide examples of %rare 
situations that might be overlooked if explanation needs are not evaluated in a structured way. Our taxonomy can provide such a structure for eliciting explanation requirements.  

In the following, we first present related work in Section~\ref{sec:relatedwork} and then discuss general characteristics of explanation needs in Section~\ref{sec:characteristics}. We identify four main categories of situations where users seek explanations: \textit{Training, Validation, Debugging} and \textit{Interaction}. In this paper, we focus on the category \textit{Interaction} and provide a taxonomy of explanation needs in interactive situations and examples in Section~\ref{sec:ExCase}.
We provide our ideas for future work on explanation cases in Section~\ref{sec:conclusion}.

%next arguments:
%\begin{itemize}
%    \item consensus of "what needs explanations (referred as  explanation cases in our paper)" is very abstract (like when  users don't meet their expectation and ..) which is not enough and we want to characterize them in more detail
   
%    \item Why are these cases useful? Benchmarking/ reference points, shared understanding, classification
    
%    \item covered domain: CPS

%\end{itemize}

%Ideas discussed in the last meeting:
%\begin{itemize}
%    \item To analyze some examples where the solutions/situations that need explanation are not very obvious and drive some requirements 
    
%    \item To state a number of scenarios, identify some attributes and characteristics of those scenarios in order to model when/what needs explanation
    
%    \item The characterization of the situation may also come from the study of works of literature, but then we specify them to more concrete situations and provide a practical example/scenario 

%\end{itemize}

\section{Related work\label{sec:relatedwork}}
%As mentioned earlier, studies on the explainability of intelligent systems in the literature are composed of the practical solutions and empirical research, where the latter 
%The former approaches are towards developing frameworks, tools, and techniques in a wide variety of domains, including but not limited to Artificial Intelligent, Robotic, Cyber-physical, Decision-making systems, etc. \cite{samek2019towards, stange2019towards, blumreiter2019towards, goud2008development}.
Empirical studies on explanations and explainable systems have been growing in various major trends. Among others, three of them are more relevant to this paper. First, works such as \cite{gregor1999explanations, collins1984action, bohlender2019towards, chari2020directions} are mainly targeting the theoretical basis and foundations of explanation. Such epistemic studies try to model and formalize the core concepts of explainability rooted in philosophy, psychology, and mathematics. Second, a set of contributions are concentrating on validation of explainable tools and theories mainly through experimental studies and analysis of human interactions with such systems to determine the necessity, pros and cons, and future direction of the explainability domain \cite{kulesza2013too, bunt2012explanations, dhaliwal1993experimental, lim2009and}.  The third branch starts from the results of the other two categories to further investigate explainability and its related issues from a requirement engineering perspective, which covers various aspects, including the requirements of an explainable system and relations between explainability and other quality requirements of systems \cite{chazette2019end, kohl2019explainability}.

The contributions in these directions together build up the background of explanation research and constitute the foundation towards constructing an efficient, user-friendly, smart, and manageable explainable system. Our work principally falls into the first category since we attempt to categorize the types, characteristics, and dimensions of generic situations in which explanations are needed. A comprehensive insight of what needs explanation is the primary step to make any system explainable. Nevertheless, knowledge in this area is sparse. The systematic survey by Nunes~et~al.~\cite{nunes2017systematic} shows only a small number of contributions focusing on this matter.

Chari et al~\cite{chari2020directions} classified a set of questions asked by users when they need some explanation from a system and the respective types of explanation for each question. Linking the users' mental models and explainability in the form of the questions they may raise, and the style of corresponding explanation has been studied~\cite{lim2009and} as well. Lim and Dey have related intelligibility queries, including ``Inputs'', ``Outputs'', ``Certainty'', ``Why'', ``Why Not'', ``What If'' and ``When'' to a set of associated explanation types that the system could generate. Nunes and Jannach identified a more extended taxonomy that captures sixteen explanation types as well as seven explanation goals~\cite{nunes2017systematic}.

Different from our paper, the main focus of these works is understanding the semantics, goal and type of the explanation itself, which plays a pivotal role in explanation generation. Whereas here, we are tackling the same problem but from a different perspective. We aim to categorize and formalize the characteristics of situations that foster the need for an explanation including the system, the user(s), and the environment. 
In this direction, Gregor and  Benbasat  \cite{gregor1999explanations} have linked various explanation-related research to build a sound theoretical base for the explainability of knowledge-based systems. They introduced a classification of the relevant factors in creating an explanation, including the so-called triggers of explanation. In particular, they defined three generic situations as needs for learning, expectation failure, and special task requirement such as report production.  In another work \cite{doshi2017towards}, the authors have articulated the need for explanation to the state of incompleteness where users seek explanation to fill the gap in understanding. In~\cite{langer2021we},a comprehensive set of explanation desideratum of five types of stakeholders have been gathered from the literature in explainable research. However, their classifications towards identifying the underlying concerns behind the explanation's need stand at a high level and lack profound specifications compared to our work. Furthermore, they omitted to present concrete scenarios for each case. We aim to provide a collection of reference scenarios to help system developers and requirement engineers drive explanation cases for their systems.

Finally, similar to our work, Zevenbergen et al. \cite{zevenbergen2020explainability} have gathered five cases for explainability along with corresponding motivation scenarios in Machine Learning based systems. Our work, however, provides a more comprehensive set of circumstances, scenarios, and analyses in cyber-physical systems.

\section{Characteristics of Explanation Needs\label{sec:characteristics}}
To overcome the lack of consensus on what needs explanation and gather a reference collection of explanation cases, we have started from and extended the high-level state-of-the-art classifications as briefly overviewed in the last section. %Additionally, we have studied and analyzed the explanation cases which have been discretely identified and discussed in the research papers, projects, and proposed tools in the expansibility domain. Furthermore, to gather as complete as possible set of situations, we have investigated the typical use case scenarios in the cyber-physical system, IoT, and smart environments to discover the circumstances that seem more challenging and require some explanations\footnote{This paper is the first results of our ongoing research, and the proposed taxonomy is a non-exhaustive set of situations we have identified so far}}. 
Accordingly, in the top layer of our proposed taxonomy, we have gathered and grouped the well-known conceptions in literature into four circumstances that motivate users to seek for some explanations: \emph{Training}, \emph{Validation}, \emph{Debugging} and \emph{Interaction} summarized in Table \ref{table:motivations}. \emph{Training} covers the situations when users would like to know about some concepts, operations, and mechanics of the system or a component of the system mainly to learn how to operate with the system. Accordingly, the required explanations %does not necessarily need to be generated at run time since they 
are generic and user-independent expositions of functions and components of a system (e.g., tool documentation for tech-savvy users or user manual for ordinary users).

\begin{table}
\renewcommand{\arraystretch}{1.3}
    \caption{Categories of situations that need explanation}
    \label{table:motivations}
    \centering
    \begin{tabular}{@{}llll@{}}
    \toprule
    Motivation & Runtime & Generality & Target user\\
    \midrule
    Training & N & Generic & End-user  \\
    Interaction & Y & Case-specific & End-user \\
    Debugging & N & Case-specific & Expert \\
    Validation & N & Generic & Expert \\
    \bottomrule
    \end{tabular}
\end{table}

The explanation for the sake of \emph{Validation} that mainly targets experts and system developers constitutes a large portion of research on explainability, particularly in the domain of artificial intelligence and machine learning (ML). ML-based applications are opaque systems generating models (trained over some pair of input and outputs) that can predict the outputs of unseen inputs. ML algorithms, however, do not allow obtaining the reason behind such correlation of input data to the associated label. It is often argued that there is a trade-off between the performance and transparency of a model~\cite{dovsilovic2018explainable}. 
Here, the need for explanation stems from the desire and necessity of opening the model’s black box to interpret the model and validate its decisions. So the explanations are required to detect any bias in the training dataset, to highlight adversarial perturbation that affects the prediction and to assure the existence of a truthful causality in the model reasoning ~\cite{arrieta2020explainable}. Accordingly, given a model, the explanations are generic since they are supposed to describe the properties and behavior of a model rather than the reason and meaning of particular outputs.

%In contrast to the \emph{Training} category, here explanations are extremely case dependent and the target audience are domain experts and model developers. 

\emph{Debugging}, which is also case-dependent, is a particular situation for particular types of systems. It occurs when a system developer wants to trace back an error or fix the system itself or a product of the system by applying some changes on firmware, mechanics, or structure of the system (e.g., code debugging in IDEs). Therefore in \emph{Debugging} cases, as well as the \emph{Validation}, the application of the explanations is in the post-mortem analysis of an executed program -either successfully or not-, hence the derivation and building process of such explanations do not happen at runtime.

It is worth highlighting that we distinguish this situation with a state where a system behaves anomalously either due to mistakes of users (e.g., wrong inputs) or lack of necessary resources (e.g., internet, permissions). There is another similar but yet distinct circumstance where a system reports some malfunctioning to inform the users although the user cannot take any action to fix the problem (e.g., when the touch screen of the system is not responsive). We have categorized the situations mentioned above under the \emph{Interaction} motivation because the trigger for the required explanation is an anomaly that damages the trust and obstructs the regular communication of the user with the system. The user needs an explanation to understand the situation and address the issue so s/he can continue to interact with the system at that particular moment rather than repairing (or building) a system.

Finally, the \emph{Interaction} motivation covers the situations where some intelligent behaviors of the system are not understandable, desirable, or expected by the user \cite{gregor1999explanations}.  
Hence, to steer the human-system interaction, the system should provide some explanations to clarify (e.g., the reason behind some actions or decisions), instruct (e.g., the required input or steps by users to proceed), and/or convince (e.g., that a decision or suggestion is valid, relevant and useful) the user \cite{dhaliwal1993experimental, hayes1983steps}. Like the \emph{Training} motivation, the primary recipients of the \emph{Interaction} category's explanations are end-users engaging with the system. However, here the confusion is directly linked to the course of actions made by the user, the context of the use, and the system's state. Accordingly, the required explanation must be derived per case. Furthermore, since the lack of adjusted explanation leaves the user baffled or disinterested and hinders further interactions, the explanation must be built dynamically at runtime.  

\begin{figure}[tp]
\centering
\includegraphics[width=1.0\columnwidth]{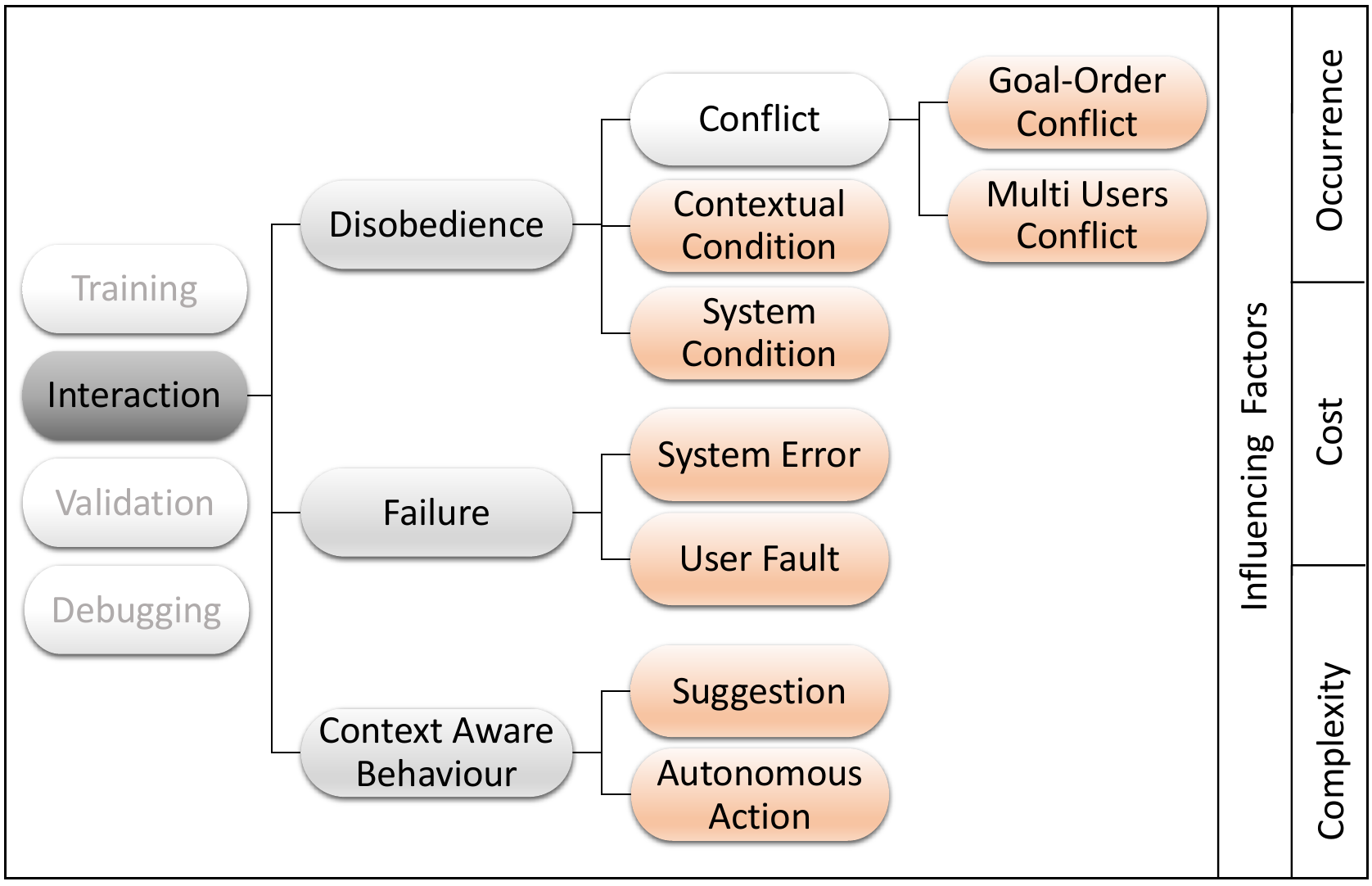}
\caption{Taxonomy of the situations that need explanation and the influencing factors}
\label{fig:taxonomy}
\end{figure}

The focus of this paper is on \emph{Interaction} motivation since it is the most relevant one to the smart environment: its recipient are the end-users, and the required explanations must be context-aware and based on the relations and communications of the user with an intelligent system. In this direction and to gather a set of situations that is as complete as possible\footnote{%This paper presents the first results of our ongoing research, and 
The proposed taxonomy is a non-exhaustive set of situations we have identified so far.} we have investigated the typical use case scenarios in cyber-physical systems, IoT, and smart environments to discover the circumstances that seem more challenging and require some explanations. Additionally, we have studied and analyzed the explanation cases which have been discretely identified and discussed in the research papers, projects, and proposed tools in the explainability domain. Overall, we have identified three sub-categories of \emph{Interaction} motivations. %They specify situations where  user and system communication becomes complicated or confusing and some provision of explanation is essential. 
First, when the system is up and running but actively ignores some direct commands of users. Second, when some failure occurs. 
Third, when a system acts intelligently based on contextual information that is more comprehensive than the user's knowledge or some complex reasons that are not very obvious/understandable  for human users. We further scrutinized such situations to discern more concrete cases and reached out to eight leaf nodes. We devoted section \ref{sec:ExCase} to elaborate on each explanation case in detail and provide a demonstration scenario for each leaf situation.

\section{Explanation Cases Taxonomy\label{sec:ExCase}}
The proposed taxonomy, depicted in Figure~\ref{fig:taxonomy}, provides a classification for the \textit{Explanation Cases} and the \textit{influencing factors}. The former identifies and discriminates various circumstances in a typical smart space that prompt users to request some explanation from the intelligent system. In the rest of this section, we describe each explanation case along with a respective scenario (See sections \ref{subsec:disobedience} to \ref{subsec:context}).

The latter covers the generic and case agnostic attributes of such situations and the intelligent system's behaviors that determine the likelihood, type, and extent of an explanation. The \textit{influencing factors} are vital elements to make the explanation generation smarter and avoid flooding users with many unnecessary explanations. 

More concretely, for each type of situation in our proposed taxonomy, as its \emph{Complexity} or \emph{Cost} increases or the number of its \emph{Occurrence} decreases, the necessity of the explanation generation rises! For example, it is more likely for a user to seek some explanation when s/he is faced with a complex autonomous decision of a system that has been taken by analysis of user behavior and interests and through complex algorithms; Rather, more straightforward automation decisions such as increasing the humidity if the humidity threshold is below some values. Similarly, another element of wonder could be the first occurrence or the infrequency of a behavior. When users are confronted with particular behavior of a system that has not been experienced by them so far, they need more explanation than for a regularly repeated activity. Finally, the cost of an action plays a pivotal role in the necessity of explaining its cause, reasons, and benefits. For example, if a system recommends reducing the temperature of a greenhouse growing hundreds of seeds, then the users are more concerned about knowing the reason and effects of following such advice compared to a situation that the system suggests them to watch a particular movie.
%*******************************> Disobedience<*****************************%
\subsection{Disobedience\label{subsec:disobedience}}
    Most often, the situation where users attempt to command a system and the system does not respond and execute the order leaves users baffled. In many cases, however, such disobedience is indeed not simply breaking the rules, but it occurs because the system is following some other rules or goals which are not very obvious for the user at that moment, or it is incapable of accomplishing the instructions due to some constraints or system conditions. In the following, we present the more specific Disobedience cases.
%=================> Conflict  <==============%
\subsubsection{Conflict}
\paragraph{Goal-Order Conflict} When users have set up a rule that might conflict with their will at particular conditions or moments, the system may decide to prioritize the predefined goal\footnote{Such prioritization could be based on user-defined rules or some intelligent algorithm of the system.} instead of the direct command and act consequently. It hence may compel the system to avoid performing orders by the user.
\begin{ExpCase}
      \item  It is 7 pm, and there is a pile of dirty dishes in the dishwasher. Bob would like to wash them before dinner but the dishwasher does not start after Bob had turned it on. The reason is that the \textit{Smarthome Manager} has deactivated the dishwasher because Bob has set up a rule to reduce the electric consumption at peak times (usually from 6 to 9 pm).
\end{ExpCase} 

\paragraph{Multi User Conflict} In a smart environment with multiple users, such as the smart home or smart office, every user may have its own goals and preferences, which may conflict with other user’s actions at some conditions or moments. Hence, the \textit{Smarthome Manager} needs to resolve such a conflict by prioritizing one user’s goals or activities over the other one. In such a situation, the user whose command has been rejected in favor of other users’ goals might be confused.

\begin{ExpCase}
      \item  Bob wants to roll up the window blinds to brighten the living room, but the system declines because Alice is watching a movie with a projector.
\end{ExpCase}

%=================> Contextual Conditions <==============%
\subsubsection{Contextual Condition} Sometimes the system chooses to avoid following a user's command due to its contextual awareness and comprehensive knowledge. The system may anticipate that the ordered action may lead to some undesirable situation or concludes the requested operation is not the best possible course of action in compliance with some rules, goals, or user preferences. 

\begin{ExpCase}
      \item  Bob receives a notification on his phone from the washing machine saying that the washing is completed. Bob then commands Coco, his assistance robot, to collect the clothes and hang them on the rack on the balcony. Coco, however, does not move to the washing machine. Bob repeats the order again but still no response from Coco! He is wondering if something is wrong with Coco! However, Coco's behavior is due to its awareness of the high chance of rain in the next few hours!
\end{ExpCase}

%=================> System Conditions <==============%
\subsubsection{System Condition} Sometimes, the system cannot perform the user’s order because of the absence of some prerequisites that involve the user’s actions to be addressed, but the user is oblivious. In this case, an explanation is necessary to bring such demands to the attention of the users.

\begin{ExpCase}
      \item  Alice is attempting to make a coffee, but the coffee machine does not start to prepare one, which makes her annoyed. It is because the coffee machine is out of coffee beans, and it needs to be refilled.
\end{ExpCase}

%*******************************> Failure<*****************************%
\subsection{Failure\label{subsec:failure}}
The next category of naturally confusing situations is when the system is not responding correctly or is operating erroneously, either due to a system error or due to a fault of the user. Here, the user requires some explanation to understand the cause of the error and some hints to return the system back to normal conditions.
\subsubsection{System Error} 
When the system is not responding, generates a wrong output or is operating erroneously, the user requires some explanation to understand the cause of the error and some hints to return the system back to the normal conditions to achieve the desired output.
\begin{ExpCase}
      \item  Bob is in the kitchen and needs to clean up the floor. He turns on his robotic vacuum cleaner (Robo cleaner) using his smartphone application and directs it to the kitchen. But it is not arriving! Bob goes to check why and notices that the Robo cleaner is stuck and it is just spinning! It is because the bumper of the cleaner must be re-assembled.
\end{ExpCase}

\subsubsection{User Fault} 
When the system fails to perform the expected behavior or generate the sought output because the user has not been following the system specification correctly, or s/he has done something wrong! In other words, in such a situation, the system has not made any mistake, but users perceive their fault as an error of the system.  

\begin{ExpCase}
      \item  Alice is in a hurry and needs to print a document. To expedite the process, she remotely connects to the printer when she is near her office and sends the document to it via her phone. When she reaches the printer to collect the paper, she finds out the printer has been printed the document on A3 paper! She immediately sends another print command by her phone, and again the printer is printing it on A3 paper! Alice is frustrated by the erroneous functioning of the printer. However, it is happening because the printer setting on her phone is not on automatic paper tray selection, and it is following the previous configuration, which has been explicitly set to take A3 papers.
\end{ExpCase}

%*****************************> Context Aware Behaviour<***************************%
\subsection{Context Aware Behaviour\label{subsec:context}}
Under this category, we analyze the situations where the intelligent behavior of the system is the source of confusion. So, users require some explanation to interpret the particular act of the system. In modern IoT systems and smart environments, the system often has a ubiquitous and comprehensive knowledge of other linked systems, environment as well as users' actions, goals, and preferences. It enables the system to learn users' tastes and better recommend a suitable course of steps for the user to follow and to perform context-aware decisions and activities in various directions---from home automation to personalized graphical user interface generation and context-aware access control for smart devices \cite{ baresi2018tdex, baresi2018fine},  to name a few.  Though the context-awareness of the system is meant to maximize user satisfaction and comfort, the autonomous and automated nature of decision making of the system or lack of transparency of reasons behind a particular recommendation may confound the users. Subsequently, some explanation can acquaint the user with the cause, purpose, and benefit of such suggestion and behavior of the system.

%=================> Suggestion  <==============%
\subsubsection{Suggestion}
One of the well-studied circumstances in the literature is when an intelligent system recommends to a user a particular item, product, or action to perform. Innately, it should clarify the motivation, reason and advantage of opting for such a thing or pursuing such action.

\begin{ExpCase}
      \item  Alice is going to bed when she receives a notification on her smartphone from the \textit{Smarthome Manager} suggesting to close all the opened windows of the house. It makes Alice wonder, due to which circumstance the system suddenly has drawn the conclusion to advise her to close the windows. ``Is there some security threat?'' Alice whispers! The system, however, has suggested it based on the temperature drop reported by the outside thermostats and the weather forecast API’s alarm that it is going to be a cold and rainy night ahead. 
\end{ExpCase}

\subsubsection{Autonomous Action}
Another progressing trend in context-aware smart homes is towards reducing human assistance and making systems more autonomous. In this direction, for example, a self-adaptive system continuously monitors itself, the user, and the environment to learn and optimize rules which let the system autonomously adapt itself to the changing environment and user preferences and activities \cite{klos2018comprehensible}. As a result, the system gradually evolves, behaves more intelligently, and independently acts or decides upon something without waiting for users' permissions. Hence, a user is confronted with some already performed actions (or the consequence of performing such actions). Without some explanations, the activity itself, or its underlying cause and benefit might not be understandable for the user.
    
\begin{ExpCase}
      \item  Bob has invited a couple of his friends to a party in his home. Suddenly, at midnight, the system turns off a couple of lights, changes the light color of others to blue, lowers the volume of the Amazon Echo playing some music, and locks the main door! Everyone, including Bob, is surprised and wonders what happened! The system, however, just followed the newly adopted rule: to put the smart-home to ``bedtime mode'' at midnight learned from Bob's routine in the last couple of weeks.
\end{ExpCase}

All these scenarios illustrate different situations that require an explanation. While some of them are obvious, others might easily be overlooked when analyzing the explainability requirements of a smart system. With our taxonomy of explanation cases, we provide precise descriptions of the characteristics of explanation cases, increase the awareness of such situations and present a collection of example situations.

\section{Conclusions\label{sec:conclusion}}
%With the rise of autonomous and intelligent systems, users seek for explanations whenever the system behavior does not match the  user’s  expectation. Thus, explainability becomes an important requirement for such systems. However, there may be several reasons for a mismatch and situations that require an explanation can be easily overlooked. In this paper, we provide a taxonomy of explanation needs that  structures  situations that require explanations according to different reasons. For each leaf node in the  taxonomy,  we  provide a concrete situation where a software system should provide an explanation. These explanation cases illustrate the different demands for explanations. 
With the rise of autonomous and intelligent systems, users seek explanations whenever the system behavior does not match the user’s expectation. Thus, explainability becomes an essential requirement for such systems. Nevertheless, while there might be various reasons for a mismatch, the situations that require an explanation have been overlooked. This paper provides a taxonomy of explanation needs that structures circumstances that require explanations according to different reasons. For each leaf node in the taxonomy,  a concrete situation where a software system should produce an explanation has been described. These explanation cases illustrate the different demands for explanations.    

%Our taxonomy and the explanation cases can already be used to guide the requirements elicitation for explanation capabilities of interactive intelligent systems. By following the taxonomy, interaction scenarios of the system, even rare cases, can be analyzed to identify necessary explanations and the influencing factors can help to further decide on when and how to explain. 
Our taxonomy can already be used to guide the requirements elicitation for explanation capabilities of interactive intelligent systems. The explanation cases help analysis of the system's interaction scenarios--even rare cases-- to identify necessary explanations, and the influencing factors assist further assessments on when and how to explain.
It is worth noting that our categorization of the situations is perceived as a means to extract the interactive explanation situations when designing an intelligent system, rather than classification of already known cases of a developed system. Accordingly, though there might be some fuzziness to discriminate the leaf nodes, it would not diminish its applicability and usefulness.

We are convinced that our findings will also help the community aligning future research on explainable systems by providing precise descriptions of explanation needs and a benchmark of explanation cases. To foster research on explainability and to profit from interdisciplinary synergies, we created a GitHub Repo\footnote{https://github.com/mersedehSa/ExplanationCases} with our explainability cases and ask the community to participate by adding further scenarios.

In future work, we will analyze our explanation cases to identify which cases are easy to detect and which need careful analyses at design time and also at runtime to provide explanations on-demand as envisioned in~\cite{blumreiter2019towards}. We envisage that our taxonomy can also provide a basis for runtime classification of system behavior with the intent to use these classes for autonomous decisions on giving explanations as envisioned in~\cite{ziesche2021anomaly}. 
%\textbf{TODO: cite date-paper? not published yet} 

Furthermore, we like to investigate whether it is possible to provide general recommendations on the kind, format and granularity of explanations %(e.g., text and pictures, text only, level of detail of explanations) 
for each leaf node of our taxonomy. 

\bibliographystyle{./bibliography/IEEEtran}
\bibliography{./bibliography/IEEEabrv,./bibliography/biblio}

\end{document}